\documentclass[12pt]{amsart}
\usepackage{graphicx,color}
\usepackage{natbib}

\usepackage{amsfonts}

\begin{document}
\bibliographystyle{ametsoc}

\title[North Atlantic hurricane changes]{On the changes in number and intensity of North Atlantic tropical cyclones}

\maketitle

\begin{center}
{\bf William M. Briggs}  \\ \vskip .05in General Internal Medicine,
Weill Cornell Medical College \\ 525 E. 68th, Box 46,
New York, NY 10021 \\ \textit{email:} mattstat@gmail.com \\
\vskip .1in
\today
\end{center}

\newpage

\begin{abstract}
Bayesian statistical models were developed for the number of tropical cyclones and the rate at which these cyclones became hurricanes in the North Atlantic.  We find that, controlling for the cold tongue index and the North Atlantic oscillation index, there is high probability that the number of cyclones has increased in the past thirty years; but the {\it rate} at which these storms become hurricanes appears to be constant.  We also investigate storm intensity by measuring the distribution of individual storm lifetime in days, storm track length, and Emanuel's power dissiptation index.  We find little evidence that the distribution of individual storm intensity is changing through time.  Any increase in cumulative yearly storm intensity and potential destructiveness, therefore, is due to the increasing number of storms and not due to any increase in the intensity of individual storms.
\end{abstract}

\section{Introduction}
It is important to be able to statistically characterize the distribution of the number of hurricanes in the North Atlantic, especially if this number is increasing or the intensity of hurricanes is changing.  Several studies have examined this.  The most important recent paper is Emanuel \citeyearpar{Ema2005}, in which he argued that hurricanes in the North Atlantic have become more destructive over the past 30 years.  To measure potential ``destructiveness", he developed a measure called the power dissipiation index, which is a function of the cubed wind speed of a storm over its lifetime (see Sec. \ref{secMethods} for a precise definition).  In his original paper, this index was not just a measure of a {\it single} storm's intensity, but a cumulative index over {\it all} the storms during the year.  Peilke \citeyearpar{Pie2005} and Landsea \citeyearpar{Lan2005} criticized the data analysis method used to demonstrate that the index was increasing; by pointing out that the smoothing method used on the raw time series data was slightly flawed, that errors in the observations should lead to a less certain statement about increases, and that the wind speed adjustments used by Emanuel were too aggressive. 

Our observation is that the lumping together of all the storms within a year has lead to a  different interpretation of what exactly is increasing: storm number or (a function of) windspeed.  Other explanations of Emanuel's findings may be that the number of cyclones has remained (distributionally) constant, but that average storm intensity has increased.  Or it may also be that the number of cyclones has increased but that the intensity on individual storms has remained constant, or even decreased.  Other combinations are, of course, possible: both storm frequency and individual storm intensity might have increased.  We examine these scenarios below.

A first step in such an analysis was taken by Elsner and Bossak \citeyearpar{ElsBos2001}, who examined the climatology and seasonal modeling of hurricane rates using a series of Bayesian hierarchical models. Using this modern approach allows us to easily make probability statements about important parameters of storm correlates and to specify the uncertainty of future predictions.  Elsner and Jagger \citeyearpar{ElsJag2004} continued the Bayesian modelling line by controlling, in their models, for the influence of the cold tounge index and the North Atlantic oscillation index.  They found that both of these indices were well correlated with the mean hurricane number.  We also use these indicies in our models below.  Elsner et al. \citeyearpar{ElsBos2001b} investigate the relationship between ENSO and hurricane numbers.  Hoyos et al. \citeyearpar{HoyAgu2006} examine these and other factors that may contribute to increases in the mean frequency of hurricanes.

Elsner et al. \citeyearpar{ElsNiu2004} and Jewson and Penzer \citeyearpar{JewPen2006} examine whether there were shifts, or change points, in the statistical distribution of hurricane numbers.  Both groups of authors did find likely changes, namely around 1900, the mid 1940s, mid 1960s, and the mid 1990s.  These shifts may have been due to actual changes in physical mechanisms (such as large-scale shifts in the atmospheric or oceanic circulations) or they may be due to changes in measurements, though all authors agree that the changes are probably a combination of both.  We take this topic up below, but we do not seek to answer why these changes take place, or even if they are certain.  It is clear enough, however, that the data has changed in character through time.  Thus, we build our models using different ranges of data in an attempt to incorporate this uncertainty.

Our approach is different from previous analyses in two ways: (1) we hierarchically model the number of tropical storms and then the {\it chance that hurricanes arise from them}, as opposed to directly modeling the number of hurricanes (or hurricane land falls); and (2) in line with \citet{WebHol2005} we characterize the distribution of storm intensity {\it within} a given year and ask whether this distribution changes through time.  We investigate storm intensity by measuring storm lifetime in days, storm track length, and Emanuel's power dissiptation index applied to invididual storms.

We use the hurricane reanalysis database (HURDAT) Jarvinen at al. \citeyearpar{JarNeu2002} and, e.g. Landsea et al. \citeyearpar{Lan2004}.  This database contains six-hourly maximum sustained 1-minute winds at 10 m, central pressures, and position to the nearest $0.1^\circ$ latitude and longitude from all known tropical storms from 1851-2006.  A cyclone was classified as a ``hurricane" if, at any time during its lifetime, the maximum windspeed ever met or exceeded 65 knots.  Obviously, this cutoff, though historical, is somewhat aribitrary and other numbers can be used: we discuss this in greater detail below.  To investigate the realtionship of North Atlantic tropical storms with ENSO, we also use the cold tongue index (CTI) \cite{DesWal1990} and e.g. \cite{Cai2003}.  And we use the North Atlantic oscillation index (NAOI) from \cite{JonJon1997}.

Section 2 lays out the statistical models and methods that we use, Section 3 contains the main results, and Section 4 presents some ideas for future research.

\section{Methods}
\label{secMethods}
We adopt, as have many before, Bayesian statistical models.  An important advantage to these models is that we can make direct probability statetments about the results.  We are also able to create more complicated and realistic models and solve them using the same numerical strategy; namely, Gibbs sampling.  We do not go into depth about the particular methods involved in forming or solving these models, as readers are likely familiar with these methods nowadays.  There are also many excellent references available, e.g. \cite{GelCar2003}.

It is important to control for factors that are known to be related, or could cause changes in, the frequency of tropical cyclones and storm intensity. There are many such possible variables, but we choose, for ease of comparison, the same indicies as cited in the paper by Elsner and Jagger \citeyearpar{ElsJag2004}.  These factors are the cold tongue index and the North Atlantic oscillation index.  Readers are encouraged to refer to the original sources and references therein to learn about these indicies.  

\subsection{Number of storms}
Most statistical analysis focuses on the number of hurricanes or the subset of landfalling hurricanes, e.g.  \cite{ElsJag2004, ElsBos2001}.  The approach here is different.  We first model the number of tropical cyclones and then model whether or not, for any given tropical cyclone, a hurricane evolves from it.  Specifically, we do {\it not} separately model the frequency of both hurricanes and cyclones, as doing this ignores the relationship of how cyclones develop into hurricanes.

We suppose, in year $i$ of $n$ years, that the number of storms is well approximated by a Poisson distribution as in 
\begin{equation}
s_i \sim \mbox{Poisson}(\lambda_i)
\end{equation}
where $\lambda_i$ describes the mean (and variance) of the number of storms.  It is of primary interest to discover whether this parameter is changing (possibly increasing) through time, controlling for known important meterological and oceaniographic variables.  Elsner and Jagger \citeyearpar{ElsJag2004} developed this same model for the number of hurricanes (and not cyclones per se).  Here we adapt it to the number of cyclones, and add in the possibility that the parameter $\lambda$ changes in a linear fashion in time (we also, for ease of reference, adopt Elsner and Jagger's notation).   Thus, we further model $\lambda_i$ as a function of the CTI and the NAOI, and allow the possibility that $\lambda_i$ changes linearly through time.  We use the generalized linear model
\begin{equation}
\label{eq1}
\log(\lambda_i) = \beta_0 +\beta_1t + \beta_2\mbox{CTI}_i+ \beta_3\mbox{NAOI}_i+ \beta_4\mbox{CTI}_i\times\mbox{NAOI}_i
\end{equation}
The prior for $\beta$ is
\begin{equation}
\mathbf{\beta} \sim MVN(a,b^{-1})
\end{equation}
that is,  $\beta = (\beta_0, \beta_1,\dots,\beta_4)$ is distributed as multivariate Normal, and $t=1,\dots,n$.  We take noninformative values on the hyperparameter $a=(0,0,\dots,0)$.  Two priors were tried for $b^{-1}=0$, which is equivalent to the standard improper flat prior, and a general covariance matrix where $b$ was given values on the off diagonals to allow for correlations between the $beta$s; both priors gave nearly equivalent results both in this model and those below.  We report results on the standard prior (technically, $b^{-1}$ is the precision and $b^{-1}=0$ is the default ``ignorance" prior in the software, which is described in Section 3). Thus, if the posterior, for example, $\Pr(\beta_1>0|\mbox{data})$ is large then we would have confidence that the mean number of storms is increasing.

We will also be interested in two other posterior probability distributions, useful for model checking: $p(\lambda_i|\mbox{data})=\int_{-\infty}^{\infty}p(\lambda_i|\beta,\mbox{data})p(\beta|\mbox{data})d\beta$, which is the posterior of the mean number of storms in year $i$ integrating over the uncertainty in (the multidimensional) $\beta_i$, and $p(g_i|\mbox{data})=\int_0^{\infty}p(g_i|\lambda_i,\mbox{data})p(\lambda_i|\mbox{data})d\lambda_i$, where the later represents the posterior predictive density (a guess) of the number of storms $g_i$ in year $i$ integrating out the uncertainty in $\lambda_i$ represented by  $p(\lambda_i|\mbox{data})$: to be clear, $p(g_i|\mbox{data})$ is the distribution we would use if we wanted to make a forecast for $s_i$ given $t$, $CTI_i$, and $NAOI_i$, integrating over the uncertainty we have in $\beta$ and $\lambda_i$.  These two distributions are not analytic and in practice we sample from them using Monte Carlo methods, which are described below.

Once a tropical storm develops it, of course, has a chance to grow into a hurricane.  If there are $s$ tropical cyclones in a year the number of hurricanes is constrained to be between 0 and $s$.  Thus, a reasonable model for the number of hurricanes $h_i$ in year $i$ given $s_i$ is 
\begin{equation}
h_i|s_i \sim \mbox{Binomial}(s_i,\theta_i)
\end{equation}
Now, if we assume that the prior on $\theta_i$ is
\begin{equation}
\label{eq2}
\theta_i \sim \mbox{Beta}(a,b)
\end{equation}
with $a=b=1$ (which represents a flat prior) then the posterior of $\theta_i$ is
\begin{equation}
\label{eq3}
\theta_i|h_i,s_i,a,b \sim \mbox{Beta}(h_i-1,s_i-h_i-1)
\end{equation}
which makes calculations easy.  This same model could of course be adapted to define hurricanes as ``major storms" using cutoffs greater than 65 knots, say greater than 114 knots for category 4 storms.  We could then estimate how many tropical cyclones evolve to these more destructive storms (we plan to address this more completely in future work, though we make a few comments below).

It is possible, however, as with $\lambda_i$, that $\theta_i$ is dependent on CTI and NAOI and that it changes through time.  To investigate this, we adopt the following logistic regression model
\begin{equation}
\log\left(\frac{\theta_i}{1-\theta_i}\right) = \beta_0 +\beta_1t + \beta_2\mbox{CTI}_i+ \beta_3\mbox{NAOI}_i+ \beta_4\mbox{CTI}_i\times\mbox{NAOI}_i
\end{equation}
where we again let 
\begin{equation}
\mathbf{\beta} \sim MVN(a,b)
\end{equation}
where $a,b$ are as before.  And again, if, for example, $\Pr(\beta_1>0|\mbox{data})$ is large then we would have confidence that the number of storms that turn into hurricanes is increasing.  In practice, however, we first transform the posteriors by exponentiation, thus the posteriors have the natural interpretation of odds ratios for the increase (or decrease) in $theta$.  So we would instead look at $\Pr(\exp(\beta_1)>1|\mbox{data})$, and if this is large, we would have good evidence that the mean rate of cyclones converting to hurricanes is increasing, i.e the odds of converting are greater than 1.

\subsection{Measures of intensity}
It may be that the frequency of storms and hurricanes remains unchanged through time, but that other characteristics of these storms have changed.  One important characteristic is intensity.  We define three measures of intensity, in line with those defined in \citet{WebHol2005}: (1) the length $m$, in days, that a storm lives; (2) the length of the track (km) of the storm over its lifetime; and (3) the power dissipation index as derived by Emmanuel, though here we apply this to each cyclone individually.

$m$ was directly available from the HURDAT reanalysis: we approximate the number of days to the nearest six-hours.  Track length was estimated by computing the great circle distance between succesive six-hour observations of the cyclone, and summing these over the storm lifetime.  The power dissipation index (PDI) is defined by
\begin{equation}
\mbox{PDI} = \int_0^TV_{\max}^3dt
\end{equation}
where $V_{\max}^3$ is the maximum sustained wind speed at 10m, and $T$ represents the total time that the storm lived.  Practically, we approximate the PDI---up to a constant---by summing the values $(V_{\max}/100)^3$ at each six-hour observation.  The PDI is a crude measure of the strength of the potential destructiveness of a tropical storm or hurricane, as cited by Emanuel \citeyearpar{Ema2005}.  Other than this measure, we say nothing directly about storm destructiveness (in terms of money etc.).

It was found that log transforms of these variables made them much more managable in terms of statistical analysis.  Transforming them led to all giving reasonable approximations of normal distributions; thus, standard methods are readily available.

For all three of these measures, we adopt a hierarchical modelling approach because we are interested if the distribution within a year of these measures changes through time.  Let $x_{i,j}$ be any of the three measures ($m$, track length, or PDI) for storm $i$ in year $j$.  Then we suppose that 
\begin{equation}
\log{x_{i,j}} \sim N(\gamma_{i,j},\chi)
\end{equation}
where
\begin{equation}
\label{eq4}
\gamma_{i,j} = \beta_0 +\beta_1t + \beta_2\mbox{CTI}_j+ \beta_3\mbox{NAOI}_j+ \beta_4\mbox{CTI}_j\times\mbox{NAOI}_j
\end{equation}
and, as before, where
\begin{equation}
\mathbf{\beta} \sim MVN(\mu,\tau)
\end{equation}
and $\mu=0$, and $\chi, \tau$, which are assumed independent of $\mu$, have inverse $\mbox{Gamma}(0.005,0.005)$ flat priors. This model is, of course, similar to those above, except that we seek to characterize the distribution of intensity {\it within} each year and see how that distribution mean might change in a (linear) way.

\section{Results}
All computations were carried out in the R statistical system \cite{RRR} using the MCMC package \cite{MCMCpack} on a Fedora Core 6 platform.  Models were fitted using Gibbs sampling.  The first 1000 simulations were considered ``burn in" and were removed from the analysis: 10,000 additional samples were calculated after this and used to approximate the posterior distirbutions. 

Data for all the measures we use was available from 1851, but we use only data from 1900 onwards.  The data from before this date, as is well known, is suspect enough to cast suspicion on any results based on them.  It is also not clear that a strict linear model over the entire period of 1900-2006 would best fit these data as observation and instrument changes through that time have changed \cite{ElsNiu2004}.  So we adopt the practice of computing each model over three different time periods: once for the entire period 1900-2006; the second for dates between 1950-2006; and the third between 1980-2006.  These choices are somewhat arbitrary, but in line with the change-point results of \cite{ElsNiu2004,JewPen2006}.  Other choices are easily made, however, and we have found that our results are robust to changes in these exact start times.  This approach also lets us check whether a linear model for increase/decrease of the parameters through time is reasonable.  We do not investigate more complicated models, such as linear change point regression models, here.  Lastly, we do not adjust the observed wind speed in any way \cite{Pie2005,Lan2005,Ema2005}.

\subsection{Number of storms}

\begin{figure}[tb]
\includegraphics{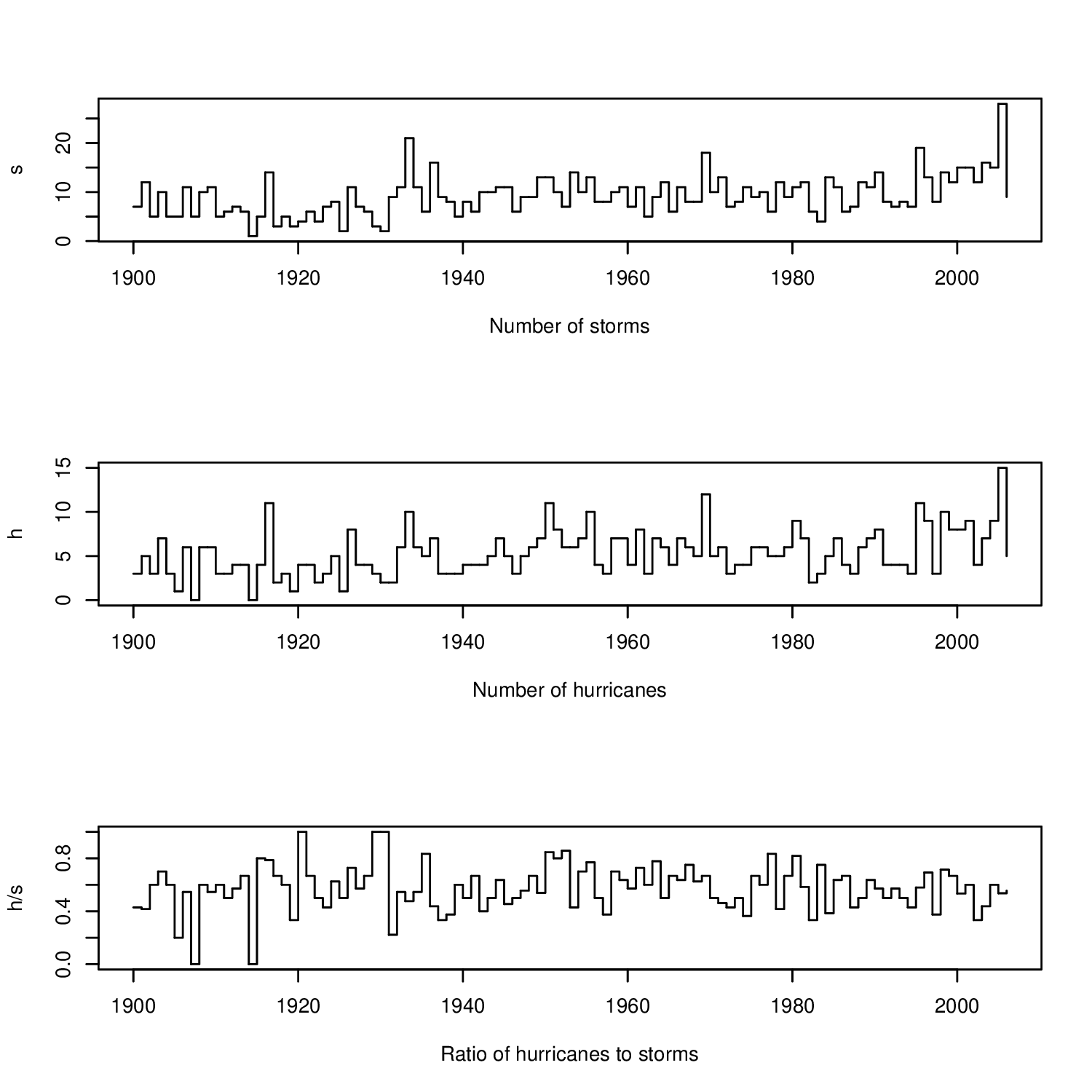}
\caption{\label{fig1} The number of storms $s$ and hurricanes $h$, and the ratio $h/s$, for the North Atlantic from 1900-2006.}
\end{figure}

The top two panels of Fig. \ref{fig1} shows the time series plots of $s$ and $h$.  There does appear, to the eye, to be an increase in $s$ in the past two decades and perhaps a smaller increase in $h$, in line with what others have found.

\begin{figure}[tb]
\includegraphics{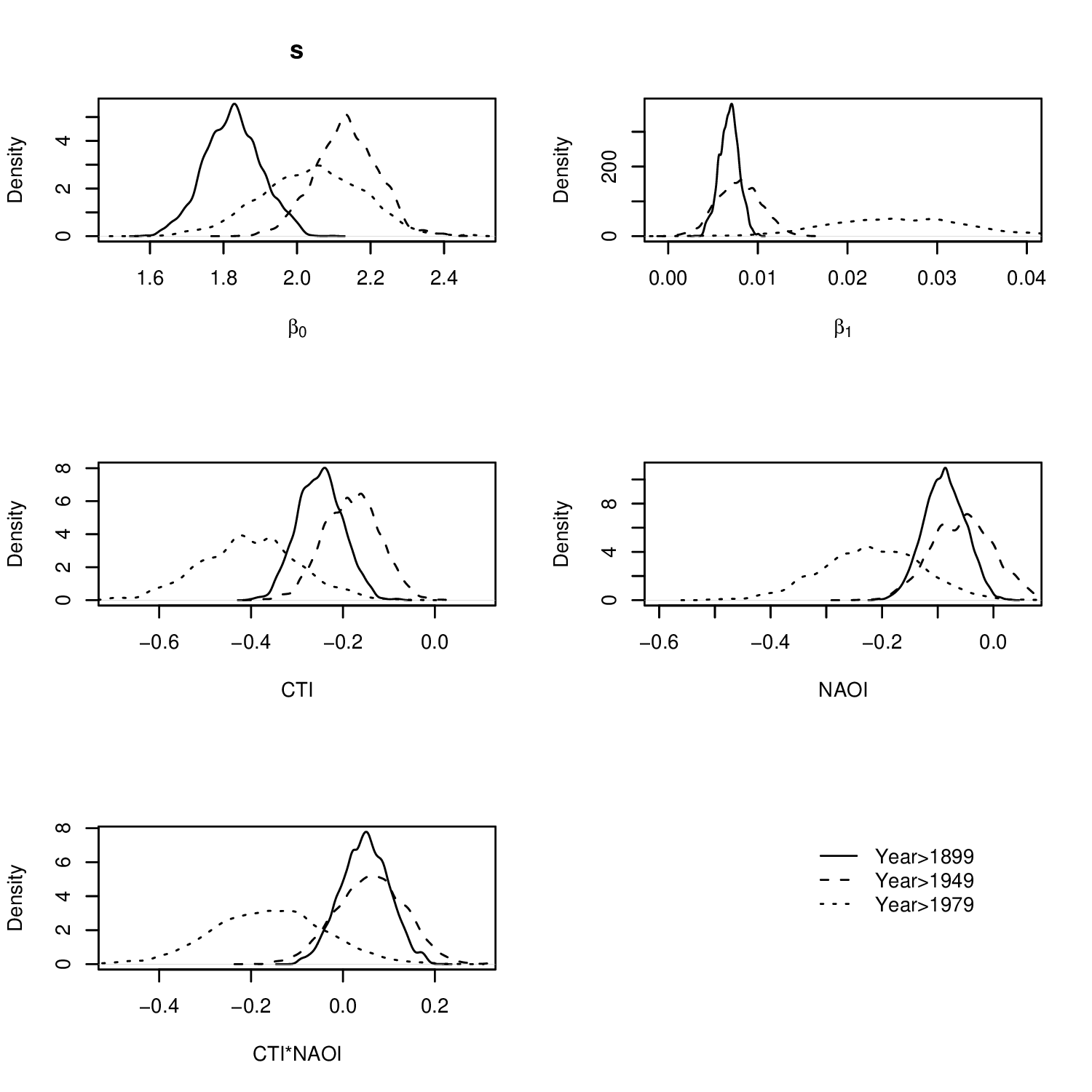}
\caption{\label{fig2} The posterior distributions for the parameters in model (\ref{eq1}).  In each case, and in all future figures, the solid line represents the model using all data from 1900-2006; the dashed line represents the model using data from 1950-2006; and the dotted line represents the model using data from 1980-2006.  Regardless of the data used, there is good evidence that $\Pr(\beta_1>0|\mbox{data})\approx 1$, which implies that the (rate of the) number of storms is increasing through time. The other parameters are discussed in the main text. Posteriors with most of their mass around 0 have little to no effect on the outcome.}
\end{figure}

Figure \ref{fig2} shows the posterior distributions from the model (\ref{eq1}).  Table 1 gives the summary statistics for this model.  In each case, and in all future figures, the solid line represents the model using all data from 1900-2006; the dashed line represents the model using data from 1950-2006; and the dotted line represents the model using data from 1980-2006.  Regardless of the data used, there is good evidence that $\Pr(\beta_1>0|\mbox{data})\approx 1$, which implies that the mean number of storms has increased through time.  However, using only the most recent (1980-2006) data does not give a very certain estimate for $\beta_1$, which can be see by noting that the probability for $\beta_1$ taking any particular value is distributed over a large range of possible values.

The parameter $\beta_0$ is the ``intercept" in model (\ref{eq1}), and is the estimate of $\lambda_i$ when the other parameters, time included, are set to 0. Thus its particular value does not directly mean much.  We can note that the intercept for $\lambda_i$ has shifted to higher values when considering the two later data sets, though we can not rule out that there has actually been change because the three distributions show great overlap.

There is also strong evidence, as Elsner and Jagger found, that the CTI ($\beta_2$) is important in estimating $\lambda_i$: greater CTIs lead to smaller $\lambda$s, and therefore to a smaller probability that the mean number of storms will be high; or more plainly, greater CTI means fewer storms.  It appears that this realtionship has strengthened in later years (1980-2006), as the distribution has shifted to smaller numbers.

Results for the NAOI ($\beta_3$) are more mixed.  Data from 1900-2006 and 1950-2006 do not give high probability that NAOI is influential.  But when only the 1980-2006 data is used, the probability is high that when NAOI is greater that $\lambda$ will be smaller, and again, a smaller probability that the number of storms will be high.  The same goes for the interaction between CTI and NAOI ($\beta_3$), though even using the 1980-2006 data does not give conclusive results.

The Table presents the same data as the figures, but only for the 1980-2006, in tabular form, so that readers can read a best (assuming absolute error loss) estimate of each $\beta$ at the median (50\%-tile).  95\% credible intervals can also be read from the table: the values from the 2.5\%- to the 97.5\%-tiles.  The last column gives the estimated probability that $P(\beta>0|\mbox{data})$.  For $\beta_1$, if this number is large, it means we have great confindence that increases are taking place.  For the other $\beta$s, if this number is near 0, it means that we are confident that $P(\beta<0|\mbox{data})\left(= 1-P(\beta>0|\mbox{data})\right)$.

\begin{table}
\begin{center}
\caption{\it Common quantiles of the model parameters and the posterior probability that these parameters are greater than 0 for  1980-2006 data only.  The exception is model (\ref{eq2}) where the probability is that the parameter is greater than 1 (because these are odds ratios).}
\label{T:standard}
{\scriptsize
\begin{tabular}{lcccc}
Parameter  &  2.5\% & 50\% & 97.5\% & $P(\beta>0|\mbox{data})$ \\ \hline\hline
\multicolumn{5}{c}{$s$} \\ \hline
$\beta_1$       &  0.011  & 0.026  & 0.041 & 0.9996    \\
CTI             & -0.61   & -0.40  & -0.19 &  $<1e-5$ \\
NAOI            & -0.40   & -0.22 & -0.059 & 0.007      \\
CTI$\times$NAOI & -0.40  & -0.16  & 0.077 &  0.095 \\

        &        & $h/s$ & & {\tiny ($P(\beta>1|\mbox{data})$)}\\ \hline
$\beta_1$       & 0.997  & 0.9999  & 1.002 & 0.47    \\
CTI             & 0.46   & 0.71  & 1.098 & 0.066 \\
NAOI            & 0.61   & 0.90 & 1.29 & 0.30      \\
CTI$\times$NAOI & 0.68  & 1.14  & 1.90 &  0.69 \\

        &        & $\mbox{Cat IV}/s$ & & {\tiny ($P(\beta>1|\mbox{data})$)}\\ \hline
$\beta_1$       &  1.0  & 1.004  & 1.009 & 0.99    \\
CTI             & 0.13   & 0.30  & .64 & 0.0001 \\
NAOI            & 0.39   & 0.84 & 1.69 & 0.33      \\
CTI$\times$NAOI & 0.19  & 0.48  & 1.20 &  0.06 \\

\multicolumn{5}{c}{$\log(m)$} \\ \hline
$\beta_1$       &  -0.006  & 0.003  & 0.011 & 0.74    \\
CTI             & -0.20   & -0.075  & 0.053 & 0.13 \\
NAOI            & -0.13   & -0.023 & 0.085 & 0.34      \\
CTI$\times$NAOI & -0.07  & 0.080  & 0.23 &  0.86 \\

\multicolumn{5}{c}{$\log(\mbox{track length})$} \\ \hline
$\beta_1$       &  -0.0073  & 0.004  & 0.012 & 0.76    \\
CTI             & -0.25   & -0.079  & 0.091 & 0.18 \\
NAOI            & -0.22   & -0.081 & 0.063 & 0.13      \\
CTI$\times$NAOI & -0.18  & 0.019  & 0.22 &  0.57 \\

\multicolumn{5}{c}{$\log(\mbox{PDI})$} \\ \hline
$\beta_1$       &  -0.012  & 0.0021  & 0.022 & 0.58   \\
CTI             & -0.65   & -0.36  & -0.056 & 0.01 \\
NAOI            & -0.34   & -0.09 & 0.16 & 0.24      \\
CTI$\times$NAOI & -0.31  & 0.043  & 0.40 &  0.59 \\
\end{tabular}
}
\end{center}
\end{table}

\begin{figure}[tb]
\includegraphics{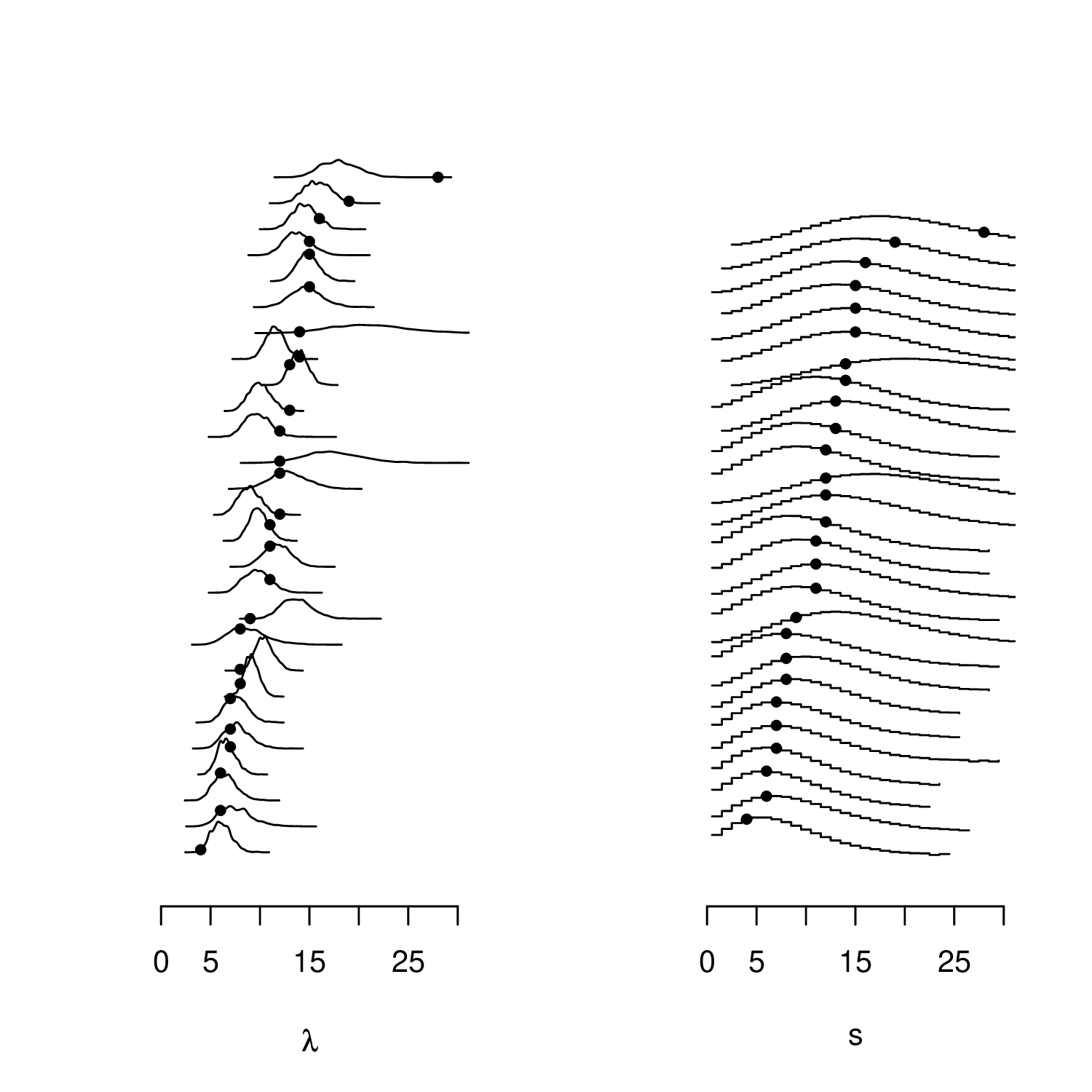}
\caption{\label{fig8} The (sorted by number of storms) posteriors $p(\lambda_i|\mbox{data})$ and $p(g_i|\mbox{data})$ for the tropical cyclones during 1980-2006.  Black dots show the actual values.  See the text for details on how these plots were constructed.}
\end{figure}

\begin{figure}[tb]
\includegraphics{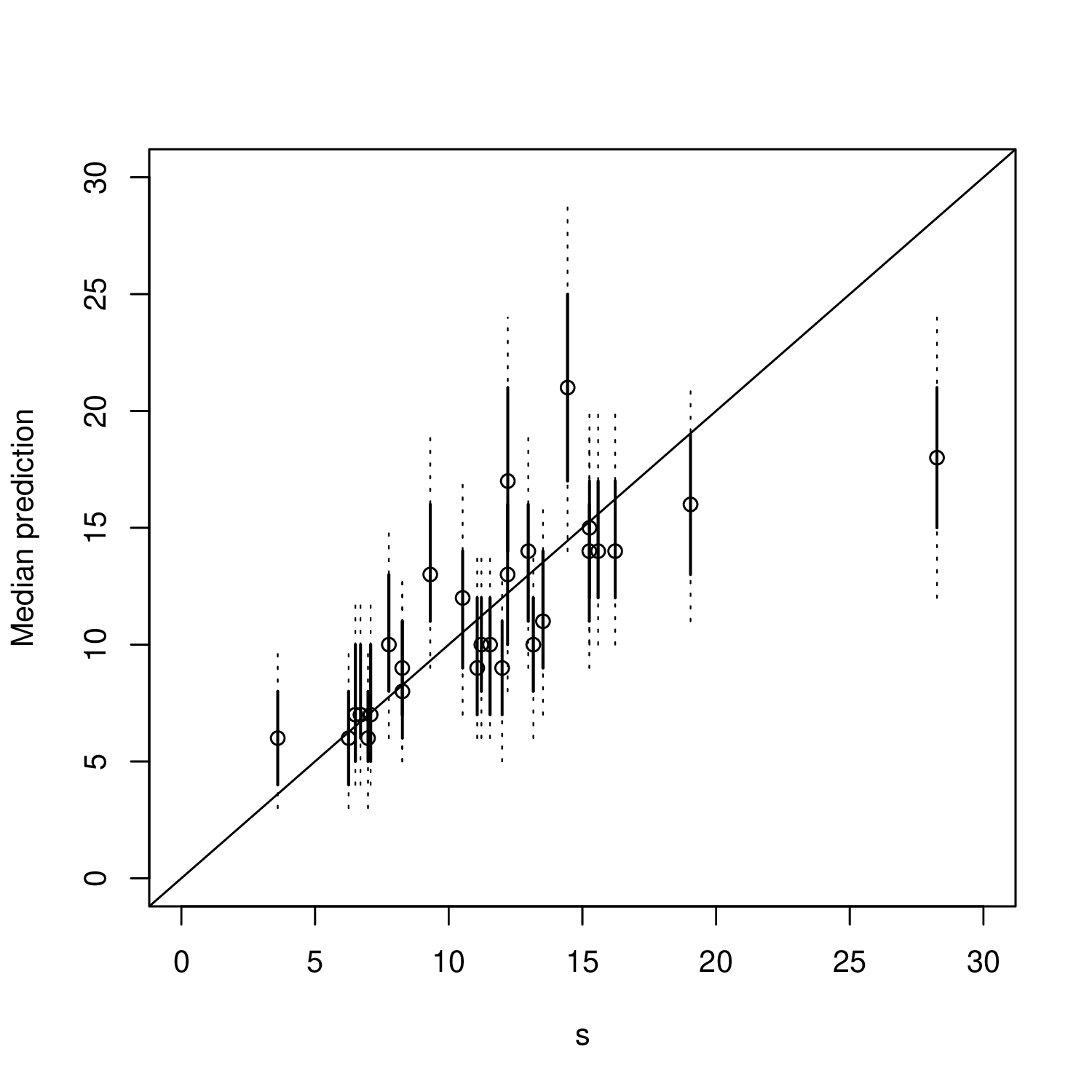}
\caption{\label{fig9} The medians (predictions) of each $g_i|\mbox{data}$ with 50\% credible intervals (solid lines) and 90\% credible intervals (thin, dotted lines), plotted by the actual values of $s_i$.  Each $s_i$ was slightly jittered to aid the eye.}
\end{figure}

These, and the following results, are only as good as the model.  To check model quality, we present Figs. \ref{fig8} and \ref{fig9}, plotted for the 1980-2006 data.  Fig. \ref{fig8} is a plot of the two posterior distributions $p(\lambda_i|\mbox{data})$ and $p(g_i|\mbox{data})$.  Overplotted on both of these distributions is the actual value (in black dots) of $s_i$.  Recall, that $p(\lambda_i|\mbox{data})$ is our best guess of the {\it mean} number of storms, and {\it not} the number of storms themselves.  This is the right-hand side of the figure: $p(g_i|\mbox{data})$.  To calculate these, we use the same MCMC results that came from the generation of the posteriors of $\beta$, and plug these into (\ref{eq1}) (along the with the appropriate values of $t$, $CTI_i$, and $NAOI_i$) and then solve for $\lambda_i$.  We first sort the actual values of $s$ and plot density estimates of the posteriors, adding an arbitrary number to each so that they may all fit on the same plot.  Fatter plots indicate more uncertainty as to the exact value of $\lambda_i$; narrower plots indicate the opposite.  Most of the dots are somewhat near the peaks of these posteriors, but certainly not all.  

The right-hand side shows the posterior of the best guess $g_i$.  These posteriors are got from, at each of the 10,000 samples of each $\lambda_i$, simulating 50 Poisson random variables with mean $\lambda_i$: then, the $50\times10,000$ values are used in building density estimates of the posterior predicitve $g_i$. Notice that these plots are jagged, which is due to the discrete nature of $g_i$ and $s_i$.  Here, to give some relief to the figure we estimated the density and took the cube root of each frequency value; which has the effect of exaggerating the high (mode) values so they are easier to see.  These posteriors are certainly more spread out, as expected.  For example, if $g\sim\mbox{Poisson}(12)$ then a 90\% prediction interval is 8 to 17.  However, it is clearer from this picture that the model is not behaving too badly.

More practically, a forecaster would pick a single number $\widehat{g}_i$ as a guess, plus a measure of uncertainty of this guess.  Fig. \ref{fig9} shows the actual values $s_i$ (slightly jittered to separate the points) by the median of $g_i|\mbox{data}$.  A one-to-one line is also shown.  Thick black lines are the 50\% credible interval, and the dotted thin lines are the 90\% credible interval.  All but 7 of the points are within the 50\% prediction intervals, and all but one (2005) are within the 90\% prediction intervals.
This gives us confidence that the model we have chosen is representative of the real ocean-atmosphere; though it is, of course, not perfect.

The bottom panel of Fig. \ref{fig1} shows the ratio $h_i/s_i$ which does not give much indication of changing through time.  Under this assumption, and using the well-known properties of the posterior (\ref{eq2}) for the data 1900-2006, we estimate that the mean fraction of converting storms is
$\widehat{\theta} = \frac{\sum h_i-1}{\sum s_i-1} = \frac{564}{992}=0.56$, and standard deviation of 0.01 (calculation not shown).  That is, once a tropical cyclone forms, there is a 56\% chance that it will evolve into a hurricane. Results for 1950-2006 are 58\%, and for 1980-2006 are 55\% (with similar standard deviations).

\begin{figure}[tb]
\includegraphics{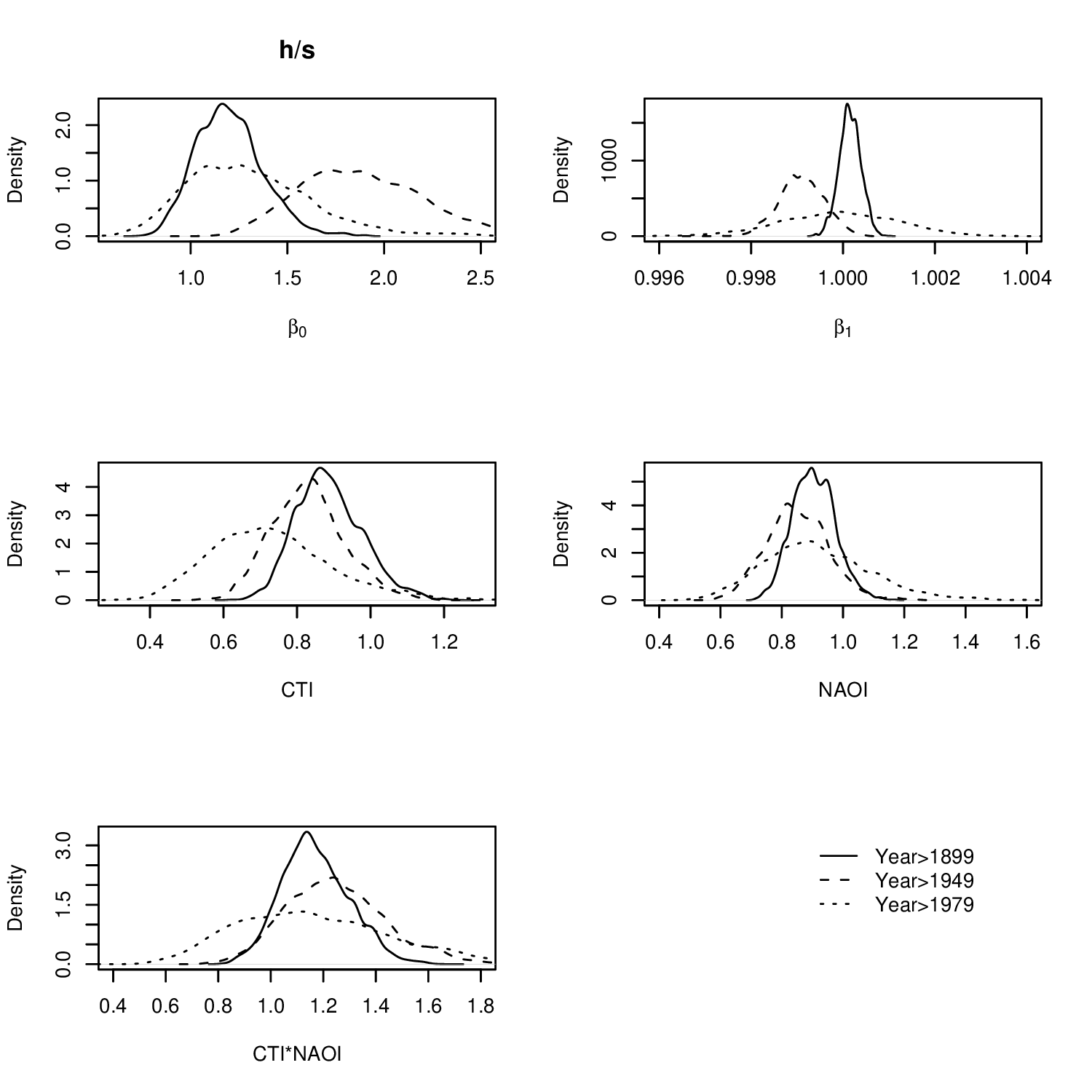}
\caption{\label{fig3} As in Fig. (\ref{fig2}) except for model (\ref{eq3}).  The results here differ in that the posteriors are in terms of odds ratios for all but the intercept, which is in odds.  Thus, posteriors with most of their mass around 1 have little to no effect on the outcome.}
\end{figure}

But applying model (\ref{eq3}) gives Fig. \ref{fig3}; summarized also in Table 1.  The rate at which storms become hurricanes does not appear to change through time, evidenced by the probability of $\beta_1$ not near 1 is small: recall that these results are in the form of odds ratios, except for $\beta_0$, which is in the form of odds.  Actually, there is some evidence, using the data from 1950-2006, that mean rate at which storms evolved was actually {\it less}, though the effect, if real, was quite small.  The best estimate, using 1980-2006, is that the odds of cyclone evolving into a hurricane are 1 per year; which is, of course, no change at all.  There is weak evidence that CTI contributes negatively, in the sense that when CTI increases, the odds of a cyclone becoming a hurricane decreases, by the odds of about 0.7 times per CTI unit.  NAOI does not appear important (most of the probability mass is around the odds ratio of 1).  The interaction of CTI and NAOI does not appear important.

The conclusion is that there is good evidence that the number of tropical cyclones has increased, but that the chance that these cyclones become hurricanes has remained constant, even after controlling for CTI and NAOI.  This also means, of course, that it is likely that the number of hurricanes has also increased.

We repeated this analysis (plots not shown, though results are in the Table) for category 4 and above hurricanes using the 1980-2006 data (defined as a cyclone which had winds meeting or exceeding 114 knots at any time during its lifetime), and find that the odds that the rate these storms evolve from tropical cyclones has increased by the odds of only 1.004 times per year (95\% credible interval 1.000 to 1.009). Much of this increase is due to 1999, when 42\% of the cyclones evolved into major storms, whereas the average rate over the remaining years is about 10\%.  Now, it is well known that standard regression models, such as this one, are quite sensitive to these sort of ``outliers", so this result should be taken with a very large grain of salt.  Too, the effect disappears when the data from 1950-2006 is used (95\% credible interval 0.998 to 1.002).  There is good evidence that CTI contributes negatively, in the sense that when CTI increases, the odds of a cyclone becoming a major storm decreases, by the odds of about 0.3 times per CTI unit.  NAOI does not appear important (most of the probability mass is around the odds ratio of 1).  The interaction of CTI and NAOI does appear important, though this is in large part driven by the CTI values.

\subsection{Measures of intensity}

\begin{center}
\begin{figure}[tb]
\scalebox{.9}{\includegraphics{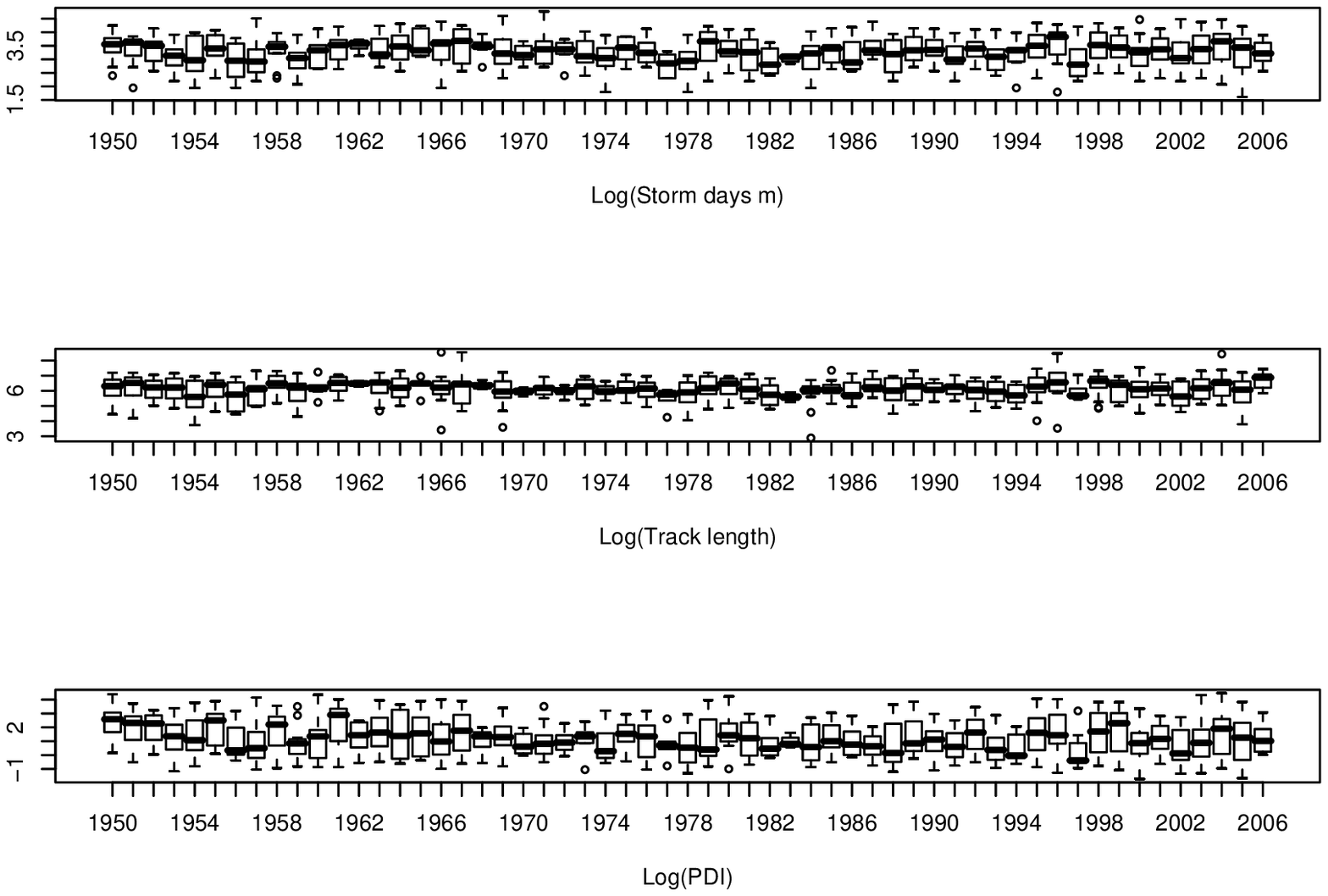}}
\caption{\label{fig4} The time series of boxplots, for each year, of $\log(m)$, $\log(\mbox{track length})$, and $\log(\mbox{PDI})$.  There is no apparent trend.}
\end{figure}
\end{center}

Figure \ref{fig4} shows the time series boxplots, for each year, of $\log(m)$, $\log(\mbox{track length})$, and $\log(\mbox{PDI})$. The boxplot gives an indication of the distribution of each measure within each year.  There is no apparent trend in the sense that, say, the medians show no systematic direction, and the other quantiles appear distributed around a central point.  If this graphical view holds under modeling, it means that cyclone distribution of intensity has not changed through time.

\begin{figure}[tb]
\includegraphics{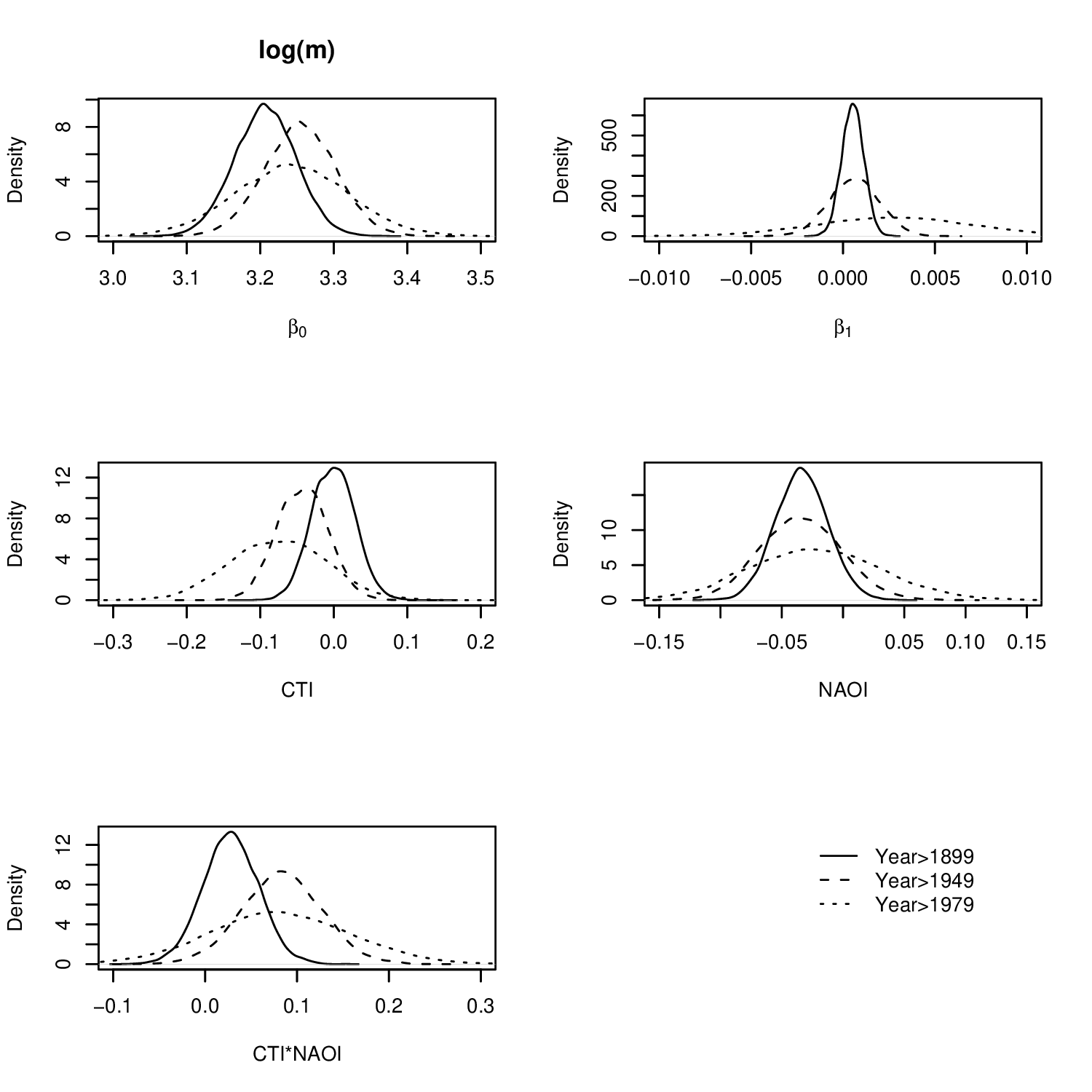}
\caption{\label{fig5} As in Fig. (\ref{fig2}) except for model (\ref{eq4}) for the logged number of storm days.}
\end{figure}

\begin{figure}[tb]
\includegraphics{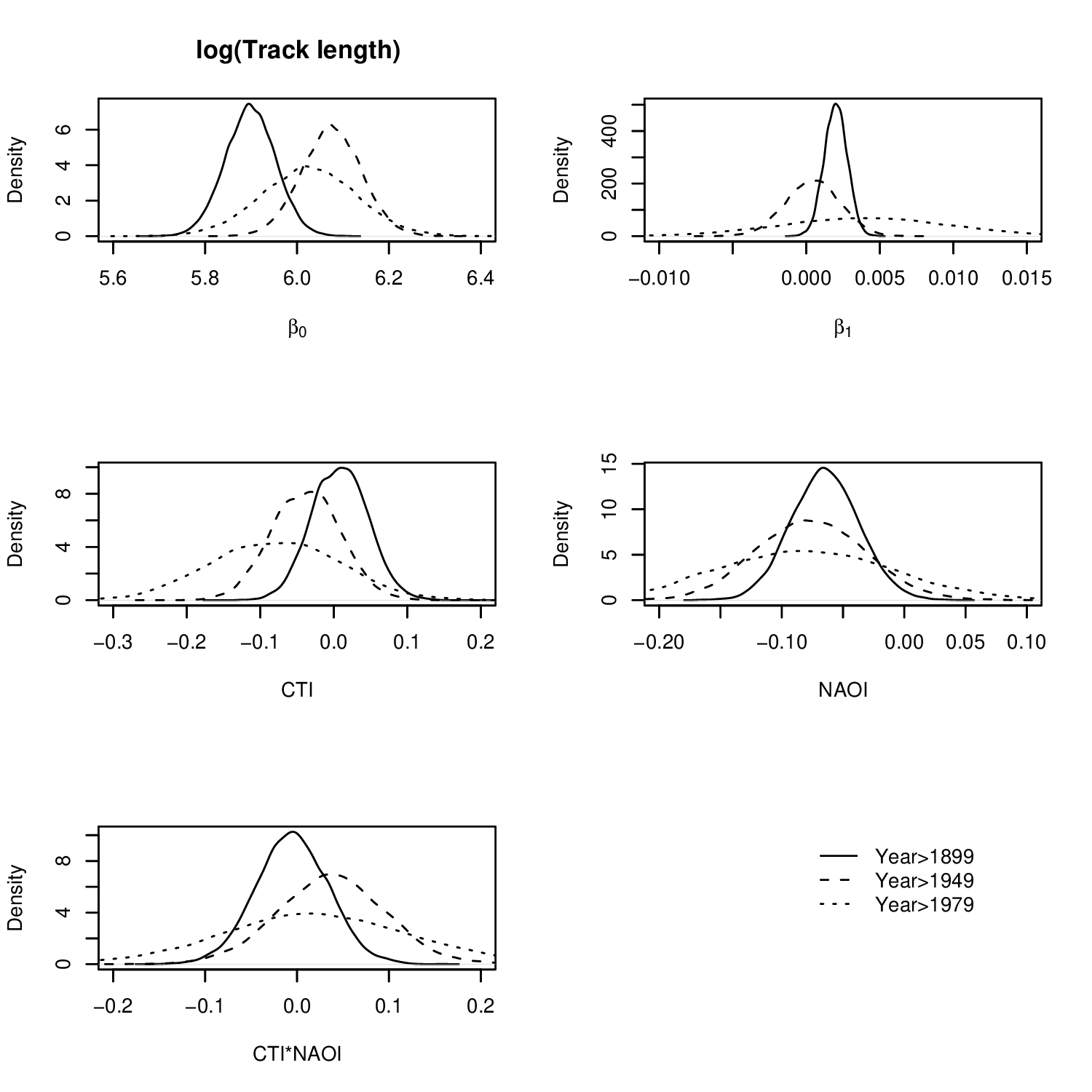}
\caption{\label{fig6} As in Fig. (\ref{fig2}) except for model (\ref{eq4}) for the logged track length.}
\end{figure}

\begin{figure}[tb]
\includegraphics{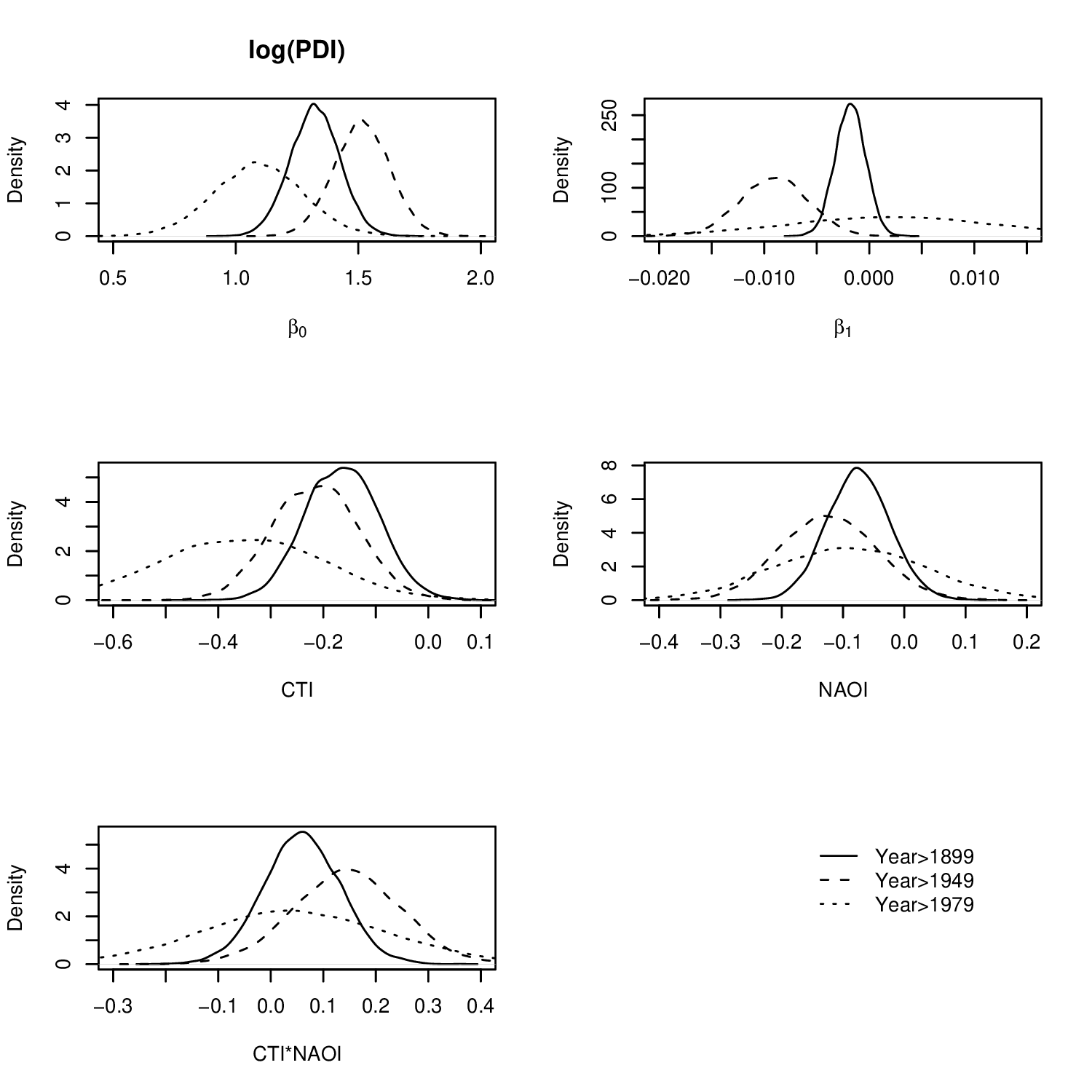}
\caption{\label{fig7} As in Fig. (\ref{fig2}) except for model (\ref{eq4}) for the logged PDI.}
\end{figure}

We now apply model (\ref{eq4}) to each of these measures.  Figs. \ref{fig5}-\ref{fig7} and Table 1 summarize the results.

For none of the measures, using any of the data time sets, does there appear to be substantial evidence that the mean of the distribution of these measures changed though time: the posteriors for $\beta_1$ in each case have most of their mass around 0.  Except perhaps for (log) PDI using the 1950-2006 data set.  Here there is good evidence that mean (log) PDI has {\it decreased} over this time period because most of the probability mass is at values less than 0. ($P(\beta<0|\mbox{data})=0.998$).  Recall that this is the distribution each individual storm's PDI within a year, and not a yearly (additive) summary.

For the (log) number of days or (log) track length, there is not much evidence that CTI or NAOI are influential, evidenced by the high probabilities of values around 0.  This is not so for (log) PDI, where there is some evidence that higher CTI and (and maybe higher NAOI) lead to smaller mean (log) PDIs.  The interaction between CTI and NAOI does not appear important.

\section{Conclusions}

We find that there is good evidence that the number of tropical cyclones in the North Atlantic basin have increased  in the past two-and-a-half decades.  This result stands even after controlling for CTI and NAOI.  These results are of course conditional on the model we used being adequate or at least it being a reasonable approximation to the data.  Diagnostics plots (Figs. \ref{fig8} and \ref{fig9}) indicate that the model performs well, though of course not perfectly, at predicting the historical data.  We make no predictions about future increases as it would be foolish to extrapolate the simple linear model we used into the future.  

We also found that the rate at which tropical cyclones become hurricanes does {\it not} appear to be changing through time, nor is it much influenced by CTI or NAOI.  There is weak evidence that the mean rate at which major (category 4 or above)  storms evolve has increased through time, though the increase is very small, and it is just as likely the model used to assess this rate is inadequate.

We find almost no evidence that the distribution of individual storm intensity, measured by storm days, track length, or individual storm PDI, has changed (increased or decreased) through time.  Any increase in storm intensity at the conglomerate yearly level, as for example found by \cite{Ema2005}, is likely due to the increased number of storms and not by the increased intensity of individual storms.  We also repeated our analysis on the distribution of each storm's (log) maximum wind speed over its lifetime and came to the same conclusion as with the other measures of intensity.

Much more exact work can be done.   A model similar to that above should certainly be used for storms across all ocean basins for which data is available, as was recently done by Webster et al. \citeyearpar{WebHol2005}.  Too, more sophisticated models could be used.  For example, spatial Bayesian models such as those developed by Wikle and Anderson \citeyearpar{WikAnd2003} for estimating tornado frequency change could be used for tropical cyclones.  This is not an easy task because tornadoes, in that model, were treated as point objects, and, of course, hurricanes vary in intensity over vast spatial lengths.  The statistical characteristics of individual tropical cyclones could be better addressed, by asking how the change in intensity (by the three measures given above, and by others such as pressure), changes through storm lifetime.  And a more complete model state-space like model for the multivariate measure of intensity (days, track length, PDI) that takes into account the correlation between these dimensions could certainly cast more light on how individual storm intensity may or may not have changed through time.

\bibliography{ams}

\end{document}